**The long mean-life-time-controlled and potentially scalable qubits composed of electric dipolar molecules based on graphene**

Yong-Yi Huang†

*MOE Key Laboratory for Nonequilibrium Synthesis and Modulation of Condensed Matter and Department of Optoelectronic Information Science and Engineering, School of Physics, Xi'an Jiaotong University, Xi'an 710049, China*

**Abstract** We propose a new kind of qubits composed of electric dipolar molecules. The electric dipolar molecules in an external uniform electric field will take simple harmonic oscillations, whose quantum states belonging to the two lowest energy levels act as the states |0>,|1> of a qubit. The qubits' excited states have a very long controlled mean life time about several hundred seconds. We can perform quantum computations by manipulating the qubits of electric dipolar molecules just like those of neutral atoms. When the qubits are used for quantum computations, the dipolar moments' orientations will harmonically oscillate along an external electric field and they will not change the directions: along or against the electric field, so the qubits can be large-scalely manufactured in graphene system. The radius of Rydberg blockade is about 100nm. The number of operated qubits reach several millions.

†Email: yyhuang@xjtu.edu.cn

## I Introduction

For a physical system to be viable in the long run, it should exhibit most or all of the elements(DiVincenzo criteria):well-defined qubits that allow for initialization into well-characterized, long-lived quantum states with the capability of high-fidelity state-dependent readout; a means to deterministically and controllably entangle individual qubits without decoherence; and the ability to transfer entanglement remotely[1]. There are several species of qubits to be used in quantum computations, for instance trapped ions[2],superconducting quantum circuitries[3], linear optics[4], semiconductor quantum dots[5,6], nitrogen vacancy centers in diamond[7] and neutral atoms[8] et al. No matter which species of qubits are manipulated, the scaling up to larger number of qubits is a central challenge. As there prevalently exist electric dipolar moments in the current quantum devices[9,10,11], the electric dipolar moments may be employed in quantum computing. De Mille proposed the novel qubits composed of ultracold diatomic molecules, whose electric dipolar moments can orient along or against an external electric field[12]. The protocol is interesting, however, the ultracold diatomic molecules in the protocol have to be cooled and trapped, which obstructs the scalability of qubits. In this paper we propose a new kind of qubits composed of electric dipolar molecules in solid. The paper is organized as follows: in section II the qubit of dipolar molecule are introduced, in section III the mean lifetime of the qubit is evaluated, in section IV Rydberg blockade and CNOT quantum gate are presented, in section V physical realization of scalable qubits is proposed based on graphene system, in section VI the conclusions are given.

## II the qubit of dipolar molecule

Given the moment of inertia of an electric dipolar molecule $J$, the electric dipolar moment $\vec{p}$, the length of dipolar molecule $l$, the an external uniform electric field strength $\vec{E}$, as shown in Fig.1, the molecule's potential in the electric field reads $V=-\vec{p}\cdot\vec{E}=-pE\cos\theta$. If the oscillation angle $\theta$ is smaller than 0.1, i.e. about $5^{\circ}$, and the potential is expanded to $\theta^4$ terms in the Taylor series, then the potential has the form of a 1D harmonic oscillator, i.e. $V=\frac{pE}{2}\left(\theta^2-\frac{\theta^4}{12}\right)$, where the constant term $-pE$ is neglected. The stationary Schrödinger's equation for the dipolar molecule's an harmonic oscillator is written as

$$\left[-\frac{\hbar^2}{2J}\frac{d^2}{d\theta^2}+\frac{pE}{2}\left(\theta^2-\frac{\theta^4}{12}\right)\right]\psi=E\psi \tag{1}$$

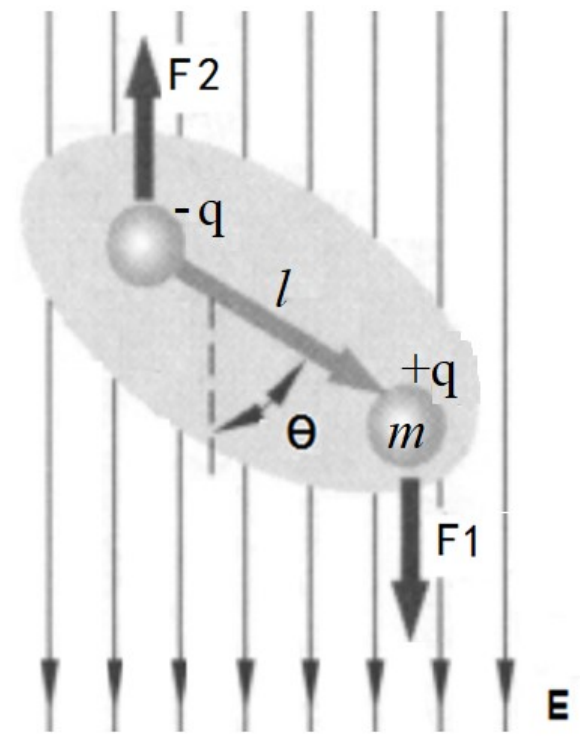

Fig. 1 The electric dipolar molecule in an external uniform electric field

Setting $x=l\theta$, the equivalent mass $m=J/l^2$, the intrinsic frequency $\omega_0=\sqrt{pE/J}$, Eq(1) becomes

$$\left(-\frac{\hbar^2}{2m}\frac{d^2}{dx^2}+\frac{m\omega_0^2}{2}x^2-\frac{m\omega_0^2}{24l^2}x^4\right)\psi(x)=E\psi(x) \tag{2}$$

To the first order approximation of the energy, Eq(2) have energy levels

$$E_n=\left(n+\frac{1}{2}-\frac{\hbar}{32J\omega_0}\left(2n^2+2n+1\right)\right)\hbar\omega_0,\ \ n=0,1,2\cdots \tag{3}$$

The nonlinearity of the energy levels of dipolar molecules resembles that of Josephson junctions in superconducting qubits, i.e. the difference of the two energy levels decreases as quantum number n increases, so the system is suitable to quantum computations. The quantum states $\psi_0$, $\psi_1$ belonging to the two lowest energy levels $E_0, E_1$ act as the states $|0\rangle, |1\rangle$ of a qubit. Without loss of generality, given the external uniform electric field strength $E=7.5V/\mu m$, the dipolar moment $p\approx 1\,Debye$, the moment of inertia $J\approx 2.5\times10^{-47}kg\cdot m^2$ [13], we obtain the intrinsic frequency $\omega_0=\sqrt{pE/J}\approx 1THz$.

**III the long controlled mean life time of the qubit**

If the qubit is in the thermal reservoir, the master equation in the interaction picture for the qubit reads

$$\frac{d\rho_{qubit}(t)}{dt}=\Gamma\left[\sigma_-\rho_{qubit}\sigma_+-\frac{1}{2}\left(\sigma_+\sigma_-\rho_{qubit}+\rho_{qubit}\sigma_+\sigma_-\right)\right] \tag{4}$$

where $\sigma_+=|1\rangle\langle 0|, \sigma_-=|0\rangle\langle 1|$. The qubit decay rate $\Gamma$ can be derived by the Weisskopf-Wigner theory, and it is equal to the Einstein's spontaneous emission coefficient[14]. The qubit decay rate is given by $\Gamma = \frac{4\omega_{eg}^3 \frac{q^2}{4\pi\varepsilon_0} \left| x_{eg} \right|^2}{3\hbar c^3}$, where $\left| x_{eg} \right|^2 = \frac{\hbar l^2 n}{2J\omega_{eg}}$ is the matrix element of a harmonic oscillator, $\hbar$ is the reduced Planck constant, *c* is the light speed in vacuum, *q* is the charge of the dipolar molecule, *n* is the quantum number, $\omega_{eg}$ is the transition frequency from an excited state $\left|1\right\rangle$ to a ground state $\left|0\right\rangle$, *l* and *J* are, respectively, the length and the moment of inertia of electric dipolar molecules. The mean life time is written as $\tau = 1/\Gamma = \frac{3Jc^3}{2\omega_{eg}^2 \frac{e^2}{4\pi\varepsilon_0} l^2 n}$ which depends on transition frequency $\omega_{eg} = \sqrt{pE/J}$ and is controlled. The mean life time increases as the transition frequency decreases. The typical parameters for dipolar molecules are $J \approx 2.5\times10^{-47} kg\cdot m^2, l = 0.13nm$ [13], $\omega_{eg} \sim 1THz, n = 1$, the mean life time $\tau$ is evaluated about $260s$, which belongs to a very long mean life time for quantum computations. The mean life time $260s$ is not only the mean life time of the excited state, but also is the coherence time of the qubit. In this paper we do not involve the depopulation time of the qubit, which should be determined by Rabi frequency.

**IV Rydberg blockade and quantum gate**

The energy level spaces of dipolar molecules in an external electric field will decrease as the quantum number *n* increases, see Eq(3), and then the qubits of electric dipolar molecules are fully equivalent to those of neutral atoms. The qubits are coupled by the electric dipole-dipole interaction, and Rydberg blockade effect entangling two qubits still works well just as the effect entangling two neutral atoms does[15]. The two dipolar molecules in Rydberg states undergo a strong dipole-dipole interaction. Consider the ground state $\left|0\right\rangle$ of a dipolar molecule coupled to its Rydberg state $\left|r\right\rangle$ with a resonant laser with a Rabi frequency $\Omega$. In the case of two dipolar molecules, the collective ground state $\left|00\right\rangle$ is still resonantly coupled to the states $\left|0r\right\rangle$ and $\left|r0\right\rangle$ containing a single Rydberg excitation. The doubly-excited state $\left|rr\right\rangle$ is shifted out of resonance by the strong van der Waals interaction U between the two molecules. In the limit $U \gg \hbar\Omega$, the double excitation is energetically forbidden(Rydberg blockade). Introducing the two collective states $\left|\psi_\pm\right\rangle = \left(\left|0r\right\rangle \pm \left|r0\right\rangle\right)/\sqrt{2}$, the collective ground state $\left|00\right\rangle$ is not coupled to $\left|\psi_-\right\rangle$ but coupled to $\left|\psi_+\right\rangle$ with the coupling $\sqrt{2}\Omega$. Starting from $\left|00\right\rangle$ and applying the laser for a duration $\pi/\left(\sqrt{2}\Omega\right)$ thus we prepares the entangled state $\left|\psi_+\right\rangle$ [16].

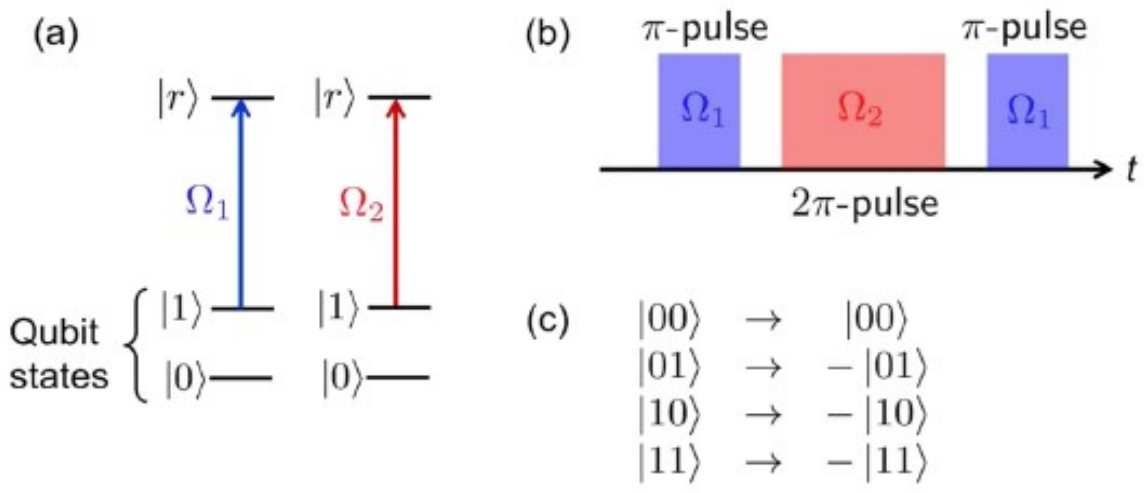

Fig.2 Principle of a two-qubit quantum gate based on Rydberg blockade.

(a) involved levels and lasers. (b) pulse sequence. (c) truth table of the phase gate[excerpted from Ref.16].

The Rydberg blockade can be used to construct fast quantum gates with dipolar molecules. As shown in Fig.2, the qubits are encoded in two lowest energy levels $|0\rangle,|1\rangle$, which can be separately addressed by lasers that couple the state $|1\rangle$ to the Rydberg state $|r\rangle$. The two molecules are close enough so that Rydberg blockade prevents the excitation $|rr\rangle$. When applying the pulse sequence shown in Fig. 2(b), if any of the qubits is initially prepared in $|1\rangle$, then the blockade makes one of the lasers off-resonant, one and only one of the molecule undergoes a $2\pi$ rotation, and the wavefunction of the system gets a minus sign at the end of the sequence. If both qubits are initially in $|0\rangle$, the laser pulses have no effect. This leads to the truth table shown in Fig. 2(c), which constructs a controlled-phase gate. The controlled-phase gate can be turned into a controlled-not(CNOT) gate using additional single-qubit gates[16]. The remarkable feature of the Rydberg gates lies in its short duration, set by the interaction energy of the two molecules. Another advantage of this protocol is that it is insensitive to the exact value of the inter-molecule interaction. From the above discussions, we know that the qubits of dipolar molecules are employed to implement quantum computing.

If the noises of room temperature 300K are suppressed, i.e. all of the qubits are in ground states in room temperature, the corresponding external electric field strength *E* is about $10^4 V/\mu m$ for the typical dipolar molecules with dipolar moment $p \approx 1\,Debye$ and the moment of inertia $J \approx 2.5\times10^{-47} kg\cdot m^2$. Actually if the condition $k_B T < \hbar\omega_0 = \hbar\sqrt{pE/J}$ is satisfied, then the noise is suppressed, we get $\omega_0 \approx 40THz$ and $E \approx 10^4 V/\mu m$. Of course, the lower the temperature is, the smaller the electric field strength is. For instrance, if the qubits work in 73K temperature, the electric field should only be about $700V/\mu m$.

**V Physical realization of scalable qubits based on graphene**

Thanks to graphene, scalability of qubits becomes probable. Fig.3 shows the nitrogen-doped graphene oxide, on which there are ‘pyridinic’ N atoms, ‘amino’ N atoms, ‘pyrrolic’ N atoms and ‘graphitic’ N atoms. As there exists charge density in graphene oxide[17], if the nitrogen-doped graphene oxide in an electric field, the ‘amino’ N atoms acts as simple pendulums which harmonically oscillate. The lowest two energy states are employed as the states |0>, |1> of a qubit. As the qubits of electric dipolar molecules are fully equivalent to those of neutral atoms, we can evaluate the radius of Rydberg blockade. The Rydberg blockade origin of dipolar molecules is different with that of natural atoms, as the dipolar moments of molecules do not change much with their oscillator states. When the two dipolar molecules blockade shift due to their interaction is large and compared to the excitation Rabi frequency $\Omega$, the excitation of the target molecule is blocked and it picks up no phase shift. Given Rabi frequency $\Omega = 1MHz$ [18,19] and typical dipolar moment $p \approx 1\,Debye$, it follows from $\frac{p^2}{4\pi\varepsilon_0 r^3} = \hbar\Omega$ that the radius of Rydberg blockade is about 100nm. As the time of gate acting on the qubits is about $1\mu s$ and the qubits possess a very long coherent time $260s$, the number of operated qubits should reach several millions. The scalability of these qubits is realized through the scalable manipulations of graphene. To be sure, our proposal is feasible in practice, as the works of some chemists have shown that dipolar systems can be adsorbed onto graphene. However, what chemists are concerned with is that the dipolar systems are switched with external fields to change the underlying electronic properties of the systems[20,21].

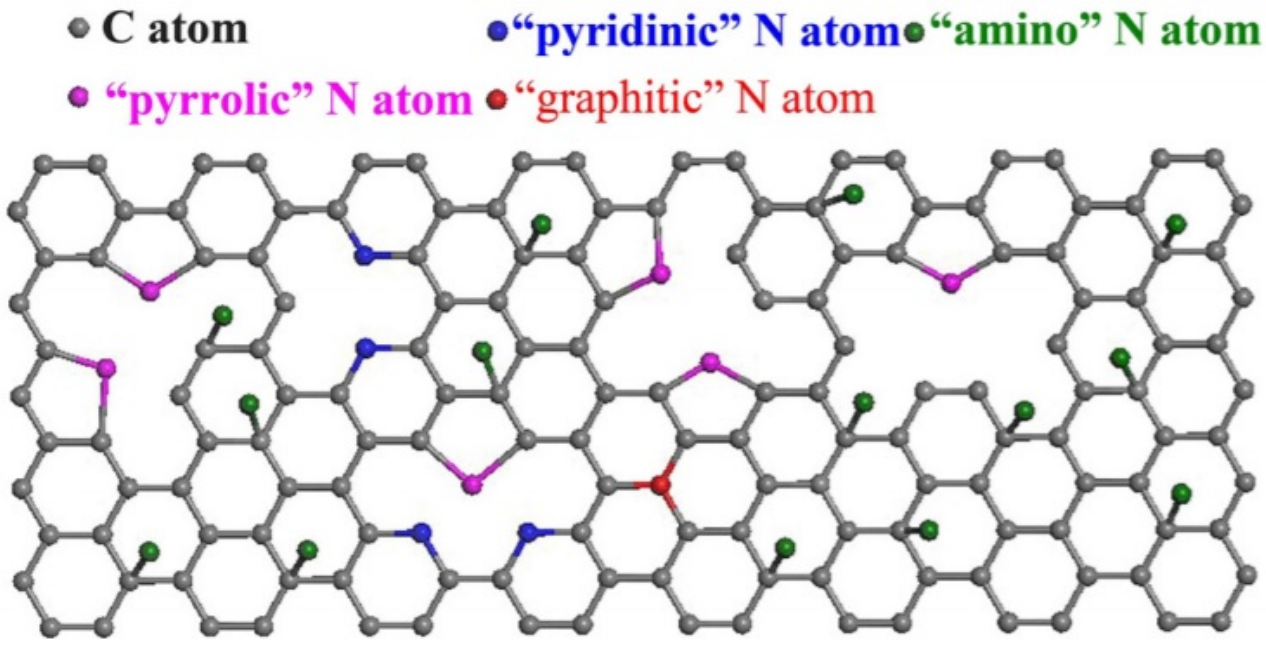

Fig.4 'Amino' N atoms of a nitrogen-doped graphene oxide[22] in an electric field as simple pendulums

**VI Conclusions**

In summary, the electric dipolar molecules in an external uniform electric field are fully equivalent to neutral atoms, the new qubits can be also used for quantum computing. The qubits of dipolar molecules have several advantages compared with the qubits of neutral atoms:(1) the mean life time of dipolar molecules' excited states $\tau$ increases as the transition frequency $\omega_{eg} = \sqrt{pE/J}$ decreases, so it is controlled by changing external electric field. The mean life time $\tau$ is evaluated extremely long $260s$ for the typical parameters, which means that decoherent phenomena will not be a serious problem in quantum computing. (2) the orientations of dipolar moments do not need to change the directions: along or against the electric field, the qubits can be large-scalely manufactured in solid; (3)the qubits do not have to be trapped, the chip's structures should be very simple. It seems feasible that the qubits on graphene are massively operated, the radius of Rydberg blockade is about 100nm, the number of operated qubits reaches several millions. The nonlinear energy levels of dipolar molecule are non-degenerate, which is helpful to manipulate the qubits precisely. Although there are other physical structures with long coherent time[23,24], our proposal of qubits incorporating long coherent time and scalability is very attractive.

**References**

[1] DiVincenzo D, Fortschr. Phys.,**48**, 771 (2000).

[2] Sutherland R, Srinivas R, Burd S, Leibfried D, Wilson A, Wineland D, Allcock D, Slichter D and Libby S, New J. Phys., **21**, 033033 (2019).

[3] Yan Z, Zhang YR, Gong M, Wu Y, Zheng Y, Li S, Wang C, liang F, Lin J, Xu Y, Guo C, Sun L, Peng CZ, Xia K, Deng H, Rong H, You JQ, Nori F, Fan H, Zhu X, Pan JW, Science, **364**, 753 (2019).

[4] Scott R, Alsing P, Smith A, Fanto M, Tison C, Schneeloch J, Hach E, Phys. Rev. A,**100**, 022322 (2019).
[5] Morton J, McCamey D, Eriksson M and Lyon S, Nature, **479**,345 (2011).
[6] Zhang X, Li HO, Cao G, Xiao Ming,Guo GC and Guo GP, Nat. Sci. Rev. **6**, 32 (2019).
[7] Hopper D, Shulevitz and Bassett L, Micromachines, **9**, 437 (2018).
[8] Weiss D and Saffman M, Phys. Today, **70**, 44 (2017).
[9] Ostapenko I, Hönig G, Kindel C, Rodt S, Strittmatter A, Hoffmann A and Bimberg D, Appl. Phys. Lett, **97**, 063103 (2010).
[10] Mar J, Baumberg J, Xu XL, Irvine A and Williams D, Phys. Rev. B, **95**, 201304(R) (2017).
[11] Groß H, Hamm J, Tufarelli T, Hess O and Hecht B, Science Advances, **4**, eaar4906 (2018).
[12] DeMille D, Phys. Rev. Lett. **88**, 067901 (2002).
[13] Experimental data for HCl, Computation Chemistry Comparison and Benchmark DataBase, release 21(August 2020), https://cccbdb.nist.gov/exp2x.asp?casno=7647010&charge=0
[14] Scully M and Zubairy M, Quantum optics, Cambridge University Press, 1997 :p249-253.
[15] Shi XF, Phys. Rev. Appl.,**9**, 051001(2018).
[16] Browaeys A, Barredo D and Lahaye T, J. Phys. B, **49**, 152001 (2016).
[17] Palacios-Lidón E, Colchero J, Ortuno M, Colom E, Benito A, Maser W and Somoza A, ACS Materials Lett., **3**, 1826 (2021).
[18] Saffman M, Walker T and Mølmer K, Rev. Mod. Phys., **82**, 2313 (2010).
[19] Walker T and Saffman M, Phys. Rev. A,**77**, 032723 (2008).
[20] Johnson P, Huang C.S., Kim M, Safron N, Arnold M, Wong B, Gopalan P and Himpsel F, Langmuir, **30**, 2559 (2014).
[21] Huang C.S., Kim M, Wong B, Safron N, Arnold M and Gopalan P, J. Phys. Chem. C, **118**, 2077(2014).
[22] Liu F, Tang N, Tang T, Liu Y, Feng Q, Zhong W and Du Y, Appl. Phys. Lett., **103**, 123108 (2013).
[23] Stipsić P and Milivojević M, Phys. Rev. B, **101**, 165302 (2020).
[24] Stavrou V and Veropoulos G, Solid State Comm. **191**, 10 (2014).